\documentclass[12pt]{article}

\begin{document}

\author{A.I.Volokitin and  B.N.J.Persson \\
Institut f\"ur Festk\"orperforschung, Forschungszentrum \\
J\"ulich, D-52425, Germany}
\title{Comment on ``No quantum friction between uniformly moving plates''}
\maketitle

\begin{abstract}
Quite recently Philbin \textit{et al} [NJP 11 (2009) 033035]  presented new theory of the van der Waals friction at zero temperature. Contrary to the previous theory they claimed that there is no 
``quantum friction'' due to quantum fluctuation of the electromagnetic field between two parallel plates moving relative to each other. We show that this theory is incorrect.
\end{abstract}

 All bodies are surrounded by a fluctuating 
electromagnetic field due to thermal and quantum fluctuations of the current density inside the
bodies.   This fluctuating field is responsible for many
important phenomena such as the radiative heat transfer, the van der Waals
interaction and the van der Waals friction between bodies. In contrast to the van der Waals interaction, for which theory is
well established, the field of van der Waals friction is still
controversial. As an example, different authors have studied the
van der Waals friction between two flat surfaces in parallel
relative motion using different methods, and obtained results
which are in sharp contradiction to each other [1-10] 
(for recent review of the theory of the van der Waals friction see \cite{Volokitin08}).

There are two approaches to the theories of the van der Waals
interaction and the van der Waals friction. In the first approach which is due to Rytov \cite{Rytov53,Rytov67,Rytov89}
the fluctuating electromagnetic field is considered as a classical
field which can be calculated from Maxwell's equation with the
fluctuating current density as the source of the field, and with
appropriate boundary conditions. This approach was used by
Lifshitz in the theory of the van der Waals interaction
\cite{Lifshitz55} and by Volokitin and Persson for the  van der
Waals friction \cite{Volokitin99,Volokitin08}. 

In the second approach the electromagnetic field is treated in the frame of
the quantum field theory. This approach was used in Ref.\cite{Lifshitz61} to
obtain the van der Waals interaction for an arbitrary inhomogeneous medium
all parts of which are at rest and in Ref. \cite{Volokitin06} to obtain the van der Waals friction  to linear order in the sliding velocity. It was shown that both approaches
 give the same results. At present there is no the quantum field theory of the van der Waals friction valid at arbitrary velocities. Such theory can be developed using the 
quantum field theory for non-equilibrium processes.

Quite recently \cite{Philbin09} Philbin \textit{et al} presented new theory of the van der Waals friction at zero temperature. Contrary to the previous theories they claimed 
that there is no ``quantum friction'' due to quantum fluctuations of the electromagnetic field
between two parallel plates moving relative to each other. Instead of using the quantum field theory for 
non-equilibrium processes they used the quantum field theory for equilibrium system which was used before in the theory of the van der Waals interaction \cite{Lifshitz61}. 
 Philbin \textit{et al} used the following expression for the vector potential
\begin{equation}
\hat{\mathbf{A}}(\mathbf{r}, t) = \sum_{\mathbf{k}, \sigma}\left[\mathbf{A}_{\mathbf{k}, \sigma}(\mathbf{r})e^{-i\omega t}\hat{a}_{\mathbf{k}, \sigma} + 
\mathbf{A}_{\mathbf{k}, \sigma}^*(\mathbf{r})e^{i\omega t}\hat{a}_{\mathbf{k}, \sigma}^+\right], \label{one}
\end{equation}
where $\mathbf{A}_{\mathbf{k}, \sigma}$ and $\mathbf{A}_{\mathbf{k}, \sigma}^*$ are a complete set of modes and $\sigma$ labels two linearly independent polarization. Using 
Eq. (\ref{one}) they derived the equal-time vacuum correlation function for the electric field $\hat{\mathbf{E}}(\mathbf{r},t)=-\partial_t \hat{\mathbf{A}}(\mathbf{r}, t)$
\begin{equation}
\langle \hat{\mathbf{E}}(\mathbf{r},t) \otimes \hat{\mathbf{E}}(\mathbf{r}^{\prime},t)\rangle = \sum_{\mathbf{k}, \sigma}\omega_k^2\mathbf{A}_{\mathbf{k}, 
\sigma}(\mathbf{r}) \otimes \mathbf{A}_{\mathbf{k}, \sigma}^*(\mathbf{r}^{\prime}). \label{two}
\end{equation} 
As in the equilibrium theory, expression on the right side of Eq. (\ref{two}) can be expressed through the Green's function of the Maxwell equation. However expression 
(\ref{one}) is only valid for stationary state when the Hamiltonian of the system does not depend on the time. In general case the time evolution of the vector potential is 
determined by expression: $\hat{\mathbf{A}}(\mathbf{r}, t) = U^+\hat{\mathbf{A}}(\mathbf{r}, 0)U$, where $U$ is an evolution operator. For stationary state
 $U=\exp(-iH_0t/\hbar)$ and time evolution of the vector potential is reduced to Eq. (\ref{one}). In the case of the van der Waals friction the Hamiltonian is time-dependent. 
Thus in this case the correlation function will be determined not by Eq. (\ref{two}) but 
by a more complicated expression  which can be derived  using  non-equilibrium quantum field theory. For example in Ref. \cite{Volokitin06} the van der Waals friction 
was calculated using Kubo formula for the friction coefficient which can be derived in the first order of the pertubation theory. 
 In fact Philbin \textit{et al} have considered the equilibrium problem 
of the van der Waals interaction between bodies whose dielectric function depends on the velocity due to the Doppler shift.  Such an approach  to the problem of 
the van der Waals 
friction is not new.  Teodorovich \cite{Teodorovich} also 
assumed that the friction force
 can be calculated as an the ordinary van der Waals interaction taking into account only the Doppler shift. The formula for the van def Waals force between moving bodies   
obtained in Ref. \cite{Philbin09}  is also incorrect because the excitations also contribute to this force and  give an addition term \cite{Volokitin08}. 

 Philbin \textit{et al} only considered the ground state and thus neglected all excitations in the system. In particular, Eq.
(\ref{two})  is only valid
for the ground state. However without excitation there is no friction. At zero temperature the van der Waalf friction originates from processes when 
 excitations are created in each body as a result of their relative motion. 
The momentum and the frequencies of these excitations are connected by $\mathbf{k}_2 = \mathbf{k}_1 + \mathbf{q}$ and $\mathbf{q}\mathbf{v} = \omega_1 + \omega_2$, where 
$\mathbf{q}$ is the momentum transfer and $\mathbf{v}$ is the sliding velocity. Thus
it is no wonder that the authors of Ref. (\cite{Philbin09}) got zero friction.
The approach they used is not suitable for the considered problem and as a
result they got incorrect results.

The origin of the van der Waals friction is closely connected with the van
der Waals interaction. The van der Waals interaction arises when an atom or
molecule spontaneously develops an electric dipole moment due to quantum
fluctuations. The short-lived atomic polarity can induce a dipole moment in
a
neighboring atom or molecule some distance away. The same is true for
extended media, where thermal and quantum fluctuation of the current
density in one body induces a current density in other body; the
interaction
between these current densities is the origin of the van der Waals
interaction. When two bodies are in relative motion, the induced current
will lag slightly behind the fluctuating current inducing it, and this is
the origin of the van der Waals friction. Delay in the system response is
a nonadiabatic
process, which is determined by excitations in the system. Without
excitations the response of the system on the external perturbation 
will be adiabatic and will result in zero-friction.
             
\vskip 0.5cm \textbf{Acknowledgment }

A.I.V acknowledges financial support from the Russian Foundation
for Basic Research (Grant N 08-02-00141-a) and DFG.

\end{document}